\documentclass[times,dvipdfmx]{akari2017} 

\shorttitle{UIBs in NGC$\,$1097}
\shortauthors{R.~Wu, F.~Galliano and T.~Onaka}

\begin{document}

\title{Evolution of the Unidentified Infrared Bands in the Nucleus of 
       the Starburst Galaxy NGC$\,$1097} 



\correspondingauthor{Fr\'ed\'eric Galliano}
\email{frederic.galliano@cea.fr}

\author{Ronin Wu} 
\affiliation{LERMA, Observatoire de Paris, PSL Research University, CNRS, Sorbonne Univrsite, UPMC Paris 06, 92190, Meudon, France} 
\author{Fr\'ed\'eric Galliano} 
\affiliation{IRFU, CEA, Universit\'e Paris-Saclay, F-91191 Gif-sur-Yvette, 
             France} 
\affiliation{Universit\'e Paris-Diderot, AIM, Sorbonne Paris Cit\'e, CEA, 
             CNRS, F-91191 Gif-sur-Yvette, France} 
\author{Takashi Onaka} 
\affiliation{Department of Astronomy, Graduate School of Science, The University of Tokyo, 7-3-1 Hongo, Bunkyo-ku, Tokyo 113-0033, Japan} 



\begin{abstract}
We present the analysis of the Unidentified Infrared Bands (UIB)
in the starburst galaxy NGC$\,$1097.
We have combined spectral maps observed with the {\it AKARI}/IRC and 
{\it Spitzer}/IRS instruments, in order to study all of the most prominent 
UIBs, from 3 to 20$\;\mu m$.
Such a complete spectral coverage is crucial to remove the common degeneracies 
between the effects of the variations of the size distribution and of the charge state of the grains.
By studying several UIB ratios, we show evidence that the average size of the 
UIB carriers is larger in the central region than in the circumnuclear ring.
We interpret this result as the selective destruction of the smallest grains by 
the hard radiation from the central active galactic nucleus.
\end{abstract}



\setcounter{page}{1}




\section{INTRODUCTION: UNDERSTANDING INTERSTELLAR DUST EVOLUTION}
\label{sec:intro}

The knowledge of the properties of interstellar dust grains and their 
variations throughout different environments is a crucial challenge in modern 
astrophysics \citep[][for reviews]{draine03c,jones17,galliano18}.
A lack of precise understanding of these properties and their evolution 
currently hampers our ability to properly unredden UV-to-mid-IR observations.
In addition, such a knowledge could provide invaluable diagnostic tools of 
deeply embedded regions.
It could also refine the prescriptions of dust-dependent physical processes 
that are used in photodissociation models \citep[\textit{e.g.}][]{le-petit06},
such as the photoelectric heating \citep[\textit{e.g.}][]{kimura16} or the 
H$_2$ formation rate on grains \citep[\textit{e.g.}][]{bron14}.

Among the dust observables, Unidentified Infrared Bands (UIB) are very 
informative features.
In comparison, the thermal dust continuum emission constitutes a rather 
degenerate tracer, as it does not contain unambiguous information on the 
composition of its carriers.
The UIBs are emitted by vibrational modes in a hydrocarbon species.
The most consensual carriers of these bands are Polycyclic Aromatic Hydrocarbons 
\citep[PAH;][for a review]{tielens08}.
However, partially hydrogenated amorphous carbons are also serious candidates 
\citep[a-C(:H); \textit{e.g.}][]{jones17}.

Numerous studies have attempted to understand the origin of the variations of 
the UIB spectra in galaxies 
\citep[\textit{e.g.}][]{hony01,vermeij02,smith07,galliano08b,lebouteiller11b,mori12,whelan13}.
Despite several breakthroughs, these studies were mostly limited to a 
sub-sample of the UIBs, due to the spectral coverage of the observations.
In particular, the $3.3\mu m$ feature (Figure~\ref{fig:spec}; right) has not 
been as extensively studied as it could be.
As we will demonstrate it in this paper, the addition of this feature can 
remove some degeneracies in the interpretation of the process at the origin of 
the variation of the UIB spectrum.

\section{AKARI AND SPITZER MID-IR SPECTRA OF NGC$\,$1097}

  \subsection{The EMPIRE Sample}

We have initiated a study extending the common wavelength coverage to 
the full UIB spetrum, by combining {\it Spitzer}/IRS
($\lambda>5\;\mu m$) and {\it AKARI}/IRC (down to $\lambda=2\;\mu m$) spectra 
of the same regions.
As part of our EMPIRE project (Evolution of Molecular gas and Pahs in 
Interstellar REgions), we have built a catalog of 160 nearby galaxies, observed 
with {\it AKARI} IRC slit mode.
It includes M$\,$51, M$\,$82, M$\,$83, M$\,$87, Cen$\,$A, IC$\,$10, NGC$\,$253, 
among others. 
$90\,\%$ of these observations overlap with {\it Spitzer} IRS footprints.
At least 20 galaxies in our sample have been well spatially resolved by both 
{\it AKARI}/IRC and {\it Spitzer}/IRS.
$25\,\%$ of the sources also have overlap with {\it Herschel}/PACS 
observations.

  \subsection{Spectral Cubes of NGC$\,$1097}

\begin{figure}[htbp]
  \begin{tabular}{cc}
    \includegraphics[width=0.5\textwidth]{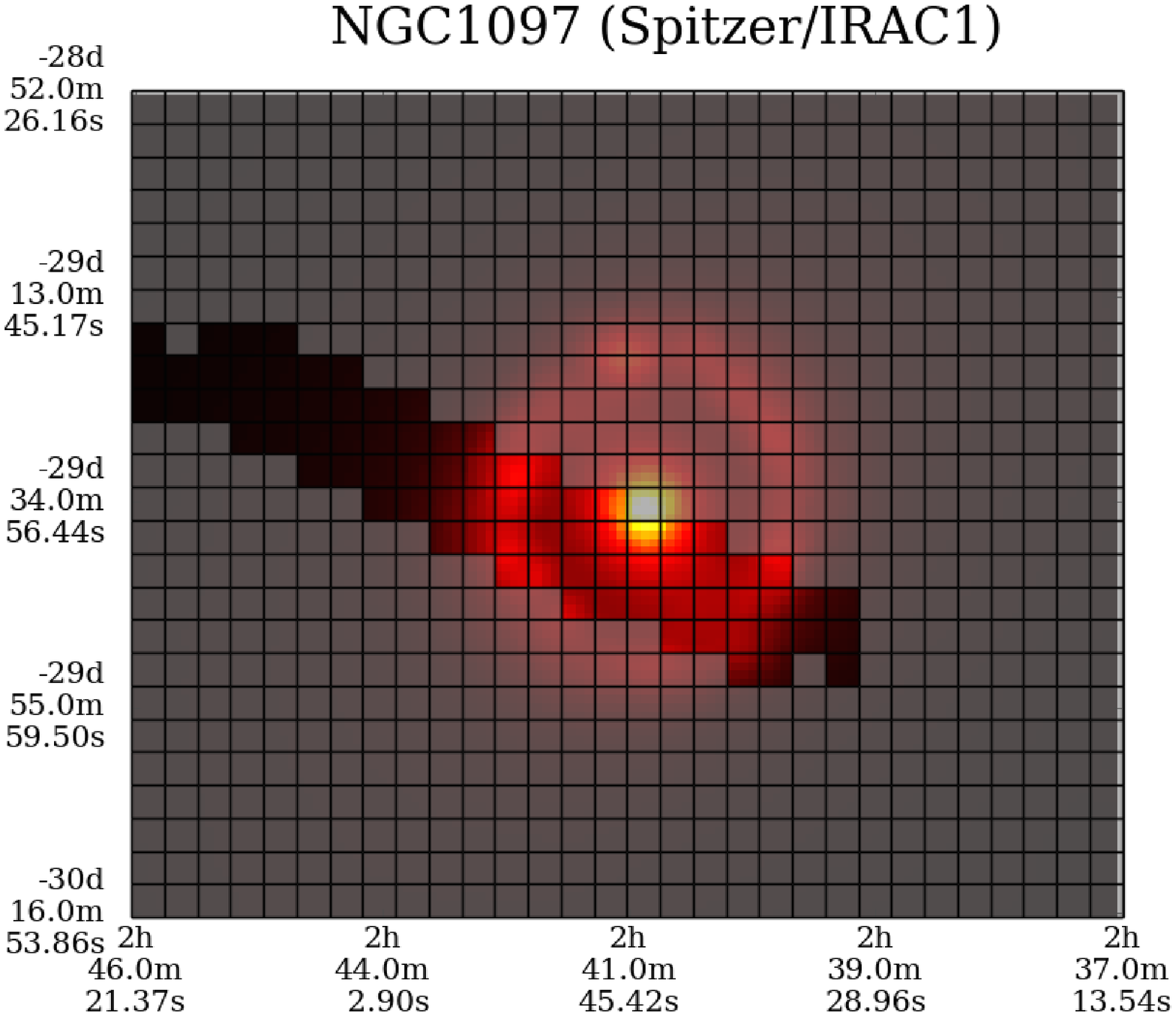} &
    \includegraphics[width=0.5\textwidth]{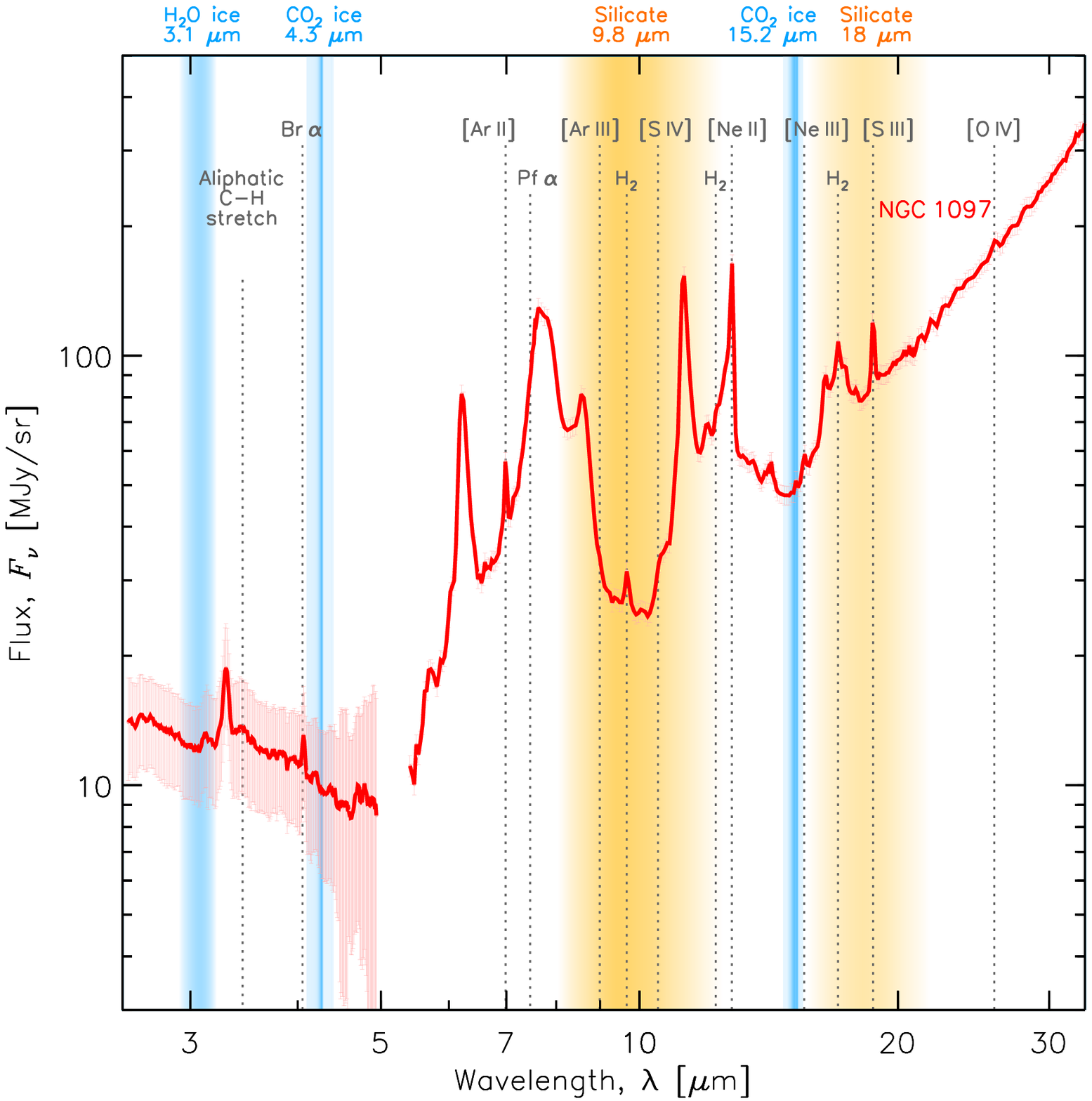} \\
  \end{tabular}
  \caption{\uline{Left:} IRAC$_{3.6\mu m}$ image of the central region of 
           NGC$\,$1097, with the footprint of the overlap of our spectroscopic 
           data.
           \uline{Right:} {\it AKARI}/IRC ($2-5\;\mu m$) 
           and {\it Spitzer}/IRS ($5-30\;\mu m$) spectra of 
           NGC$\,$1097. 
           The most notable lines are labeled in grey.
           Ice and silicate absorption features are shown in blue and orange, 
           respectively.}
  \label{fig:spec}
\end{figure}
The present paper focuses on one source of the EMPIRE catalog, the starburst 
galaxy NGC$\,$1097.
It is a Seyfert 1 galaxy of SB(s)b type with a distance of 19.1~Mpc 
\citep{willick97}.
Its mid-IR morphology (Figure~\ref{fig:spec}; left) exhibits a circumnuclear 
star forming ring, and a central active galactic nucleus (AGN).
The {\it Spitzer}/IRS spectrum of NGC$\,$1097 has been studied by 
\citet{smith07} and \citet{beirao12}, and
its {\it AKARI}/IRC spectrum by \citet{kondo12}.
The integrated mid-IR spectrum of the overalp or IRC and IRS footprints is 
shown in Figure~\ref{fig:spec} (right).
All the main UIBs, between 3 and 20~$\mu m$, are well detected, as well as 
several molecular Hydrogen and ionic lines.
The steeply rising continuum in the 20-30~$\mu m$ range, likely coming from hot 
equilibrium grains in compact H$\,${\sc ii} regions, is a sign of sustained 
star formation activity.

\section{BAND RATIO VARIATIONS AS A SIGN OF GRAIN EVOLUTION}

  \subsection{Observed Band Ratio Correlations}

We have performed a decomposition of the spectrum of each pixel, in 
order to extract the intensity of the main features.
The observed relation between select band ratios provides qualitative insights 
on the physical process driving the intensity variations.
The top-left panel of Figure~\ref{fig:correl} shows the classic 
PAH$_{7.7}$/PAH$_{11.3}$ ratio as a function of PAH$_{6.2}$/PAH$_{11.3}$.
We can see, on the bottom-right panel of Figure~\ref{fig:correl}, that if only 
the PAH ionization fraction varies, then these two ratios should be perfectly 
correlated.
Indeed, the top-left panel of Figure~\ref{fig:correl} shows strong 
correlation of these ratios, with a factor of $\simeq2$ in dynamical range.
However, this correlation plot is degenerate with other effects, such 
as the variation of the size distribution.

As stated in Section~\ref{sec:intro}, the $3.3\;\mu m$ feature is a good size 
indicator.
We can see, on the bottom-right panel of Figure~\ref{fig:correl}, that the 
$3.3\;\mu m$ to $6-9\;\mu m$ feature ratio is higher for PAH mixtures with a 
smaller average size.
The top-right panel of Figure~\ref{fig:correl} shows the same 
PAH$_{3.3}$/PAH$_{11.3}$ ratio as a function of PAH$_{7.7}$/PAH$_{3.3}$.
If the charge state was the only factor, these two ratios should be perfectly 
anticorrelated, which is not the case.

  \subsection{Interpretation in Terms on Environmental Variations}

\begin{figure}[htbp]
  \begin{tabular}{cc}
    \includegraphics[width=0.5\textwidth]{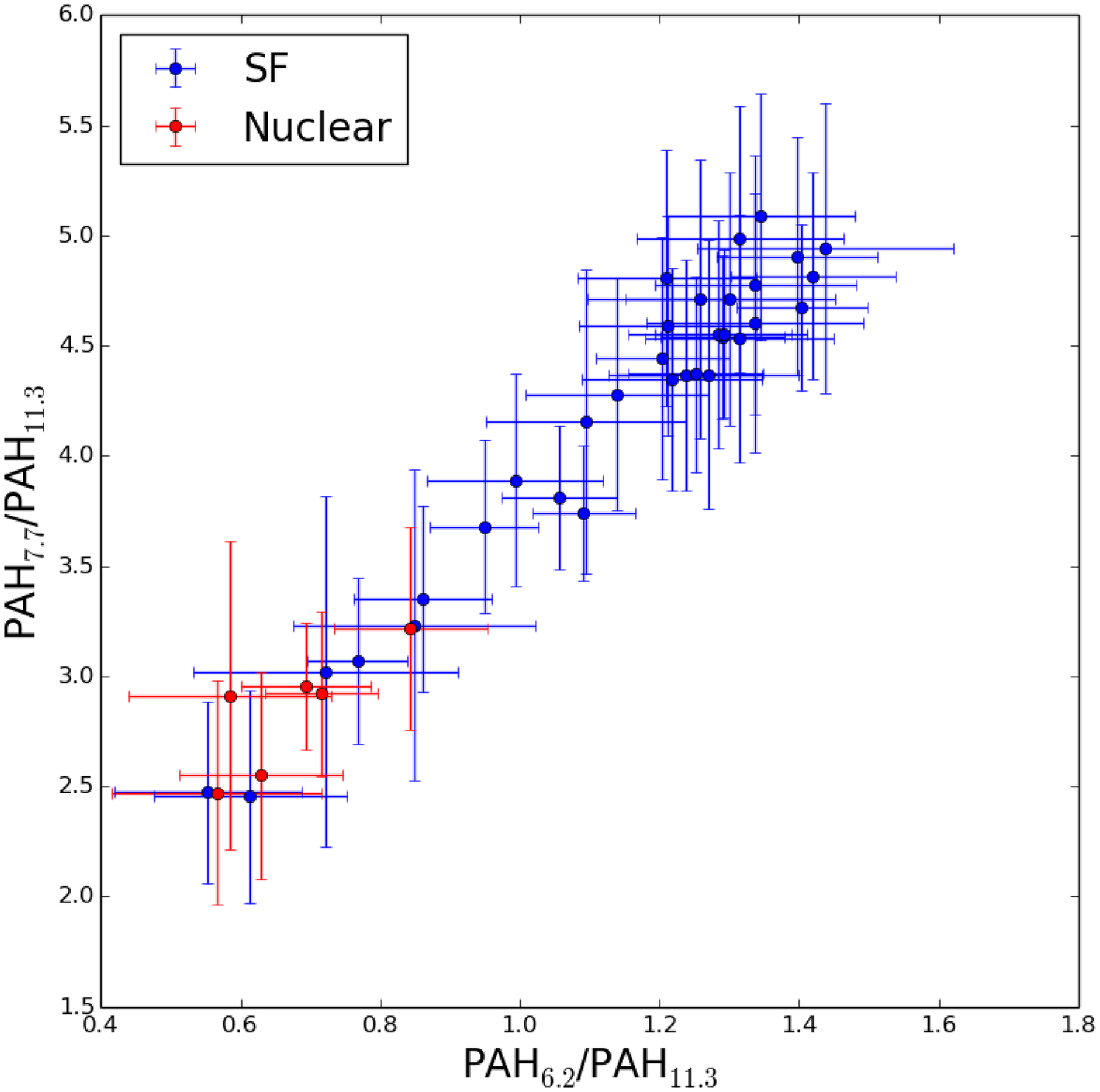} &
    \includegraphics[width=0.5\textwidth]{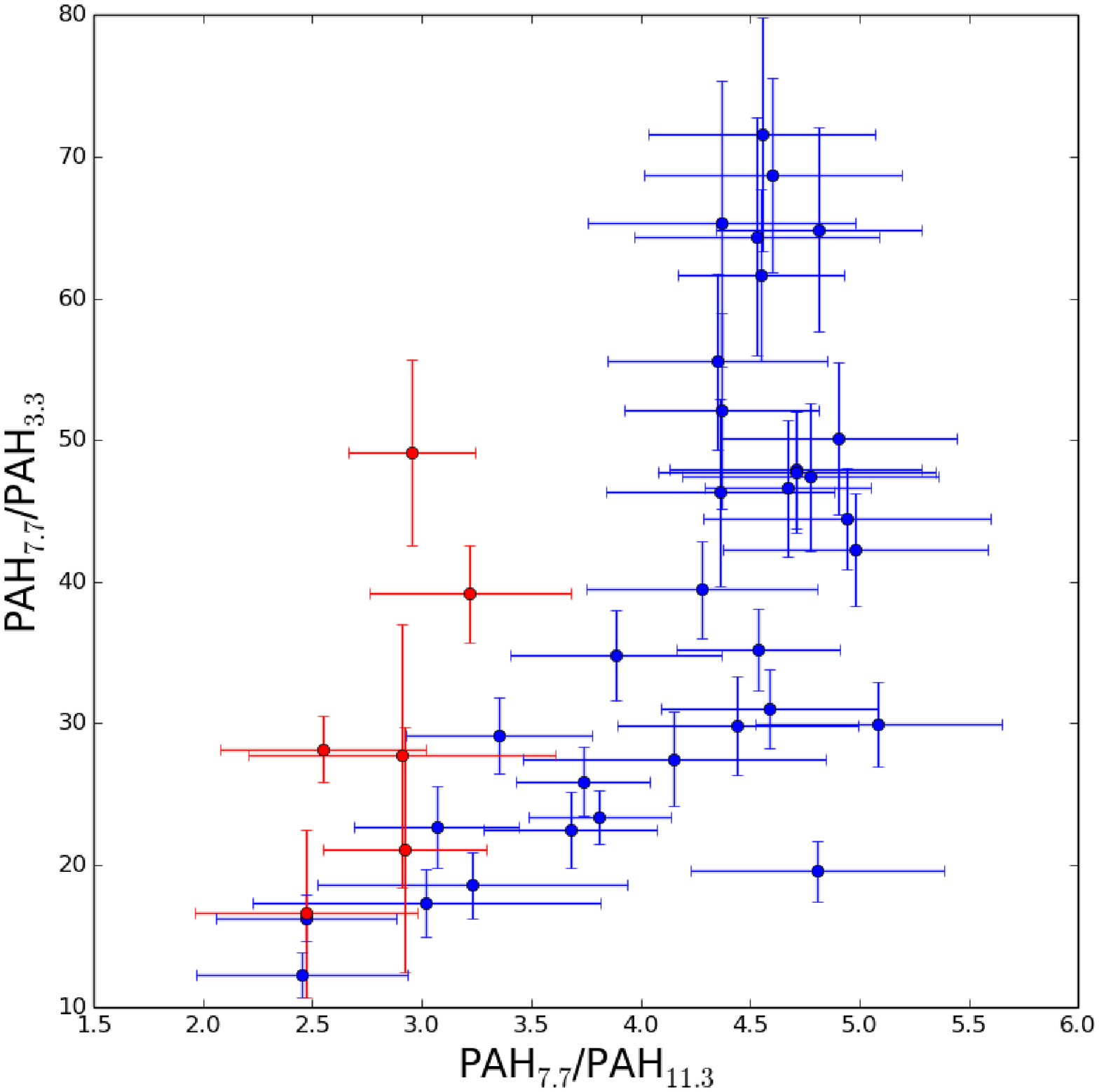} \\
    \includegraphics[width=0.5\textwidth]{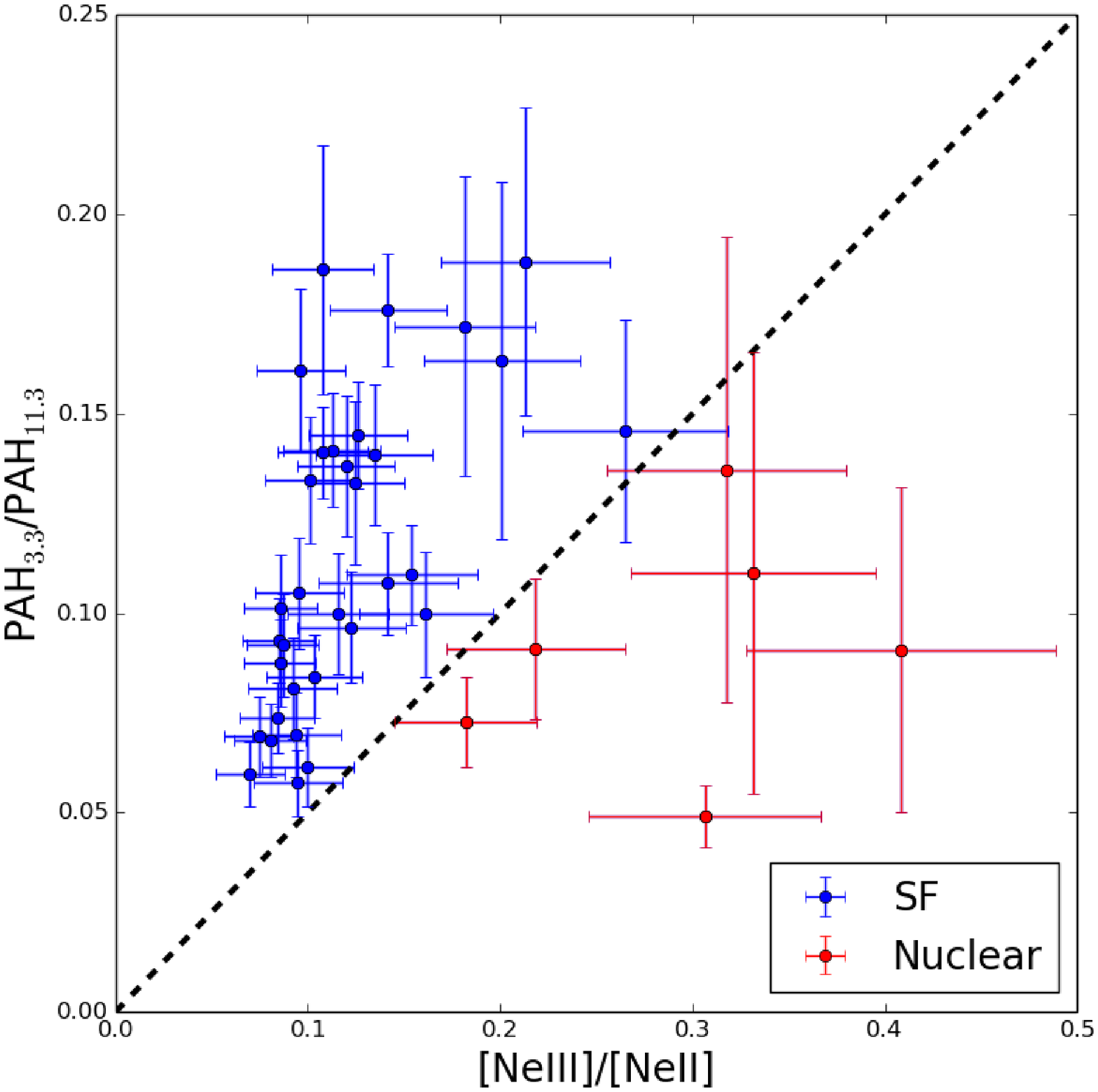} &
    \includegraphics[width=0.5\textwidth]{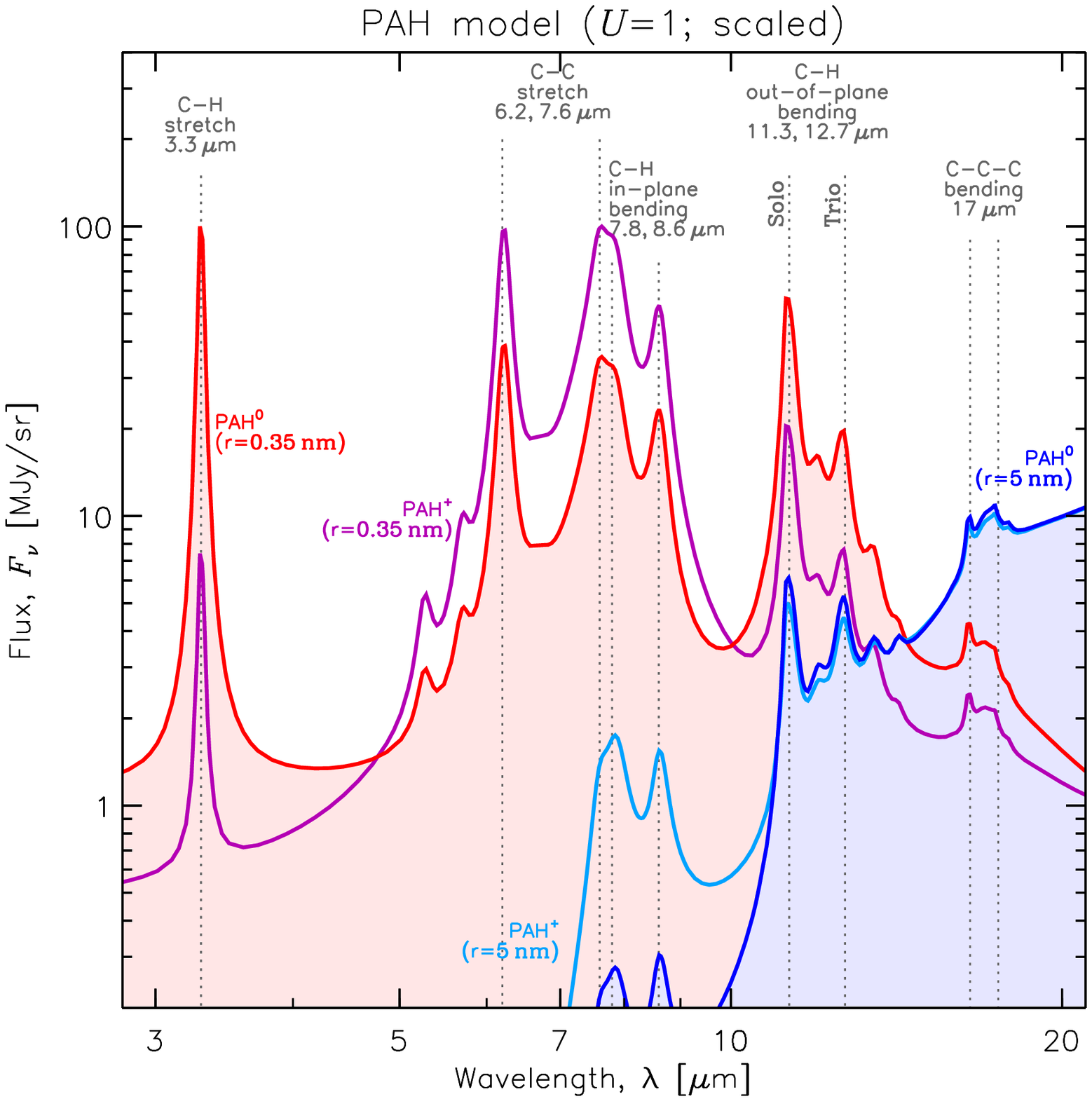} \\
  \end{tabular}
  \caption{\uline{Top and bottom-left:} pixel by pixel correlations between 
           intensity ratios.
           PAH$_\lambda$ is the intensity of the PAH feature at 
           $\lambda\;\mu m$.
           Due to the high signal-to-noise ratio, the {\it Spitzer} band ratio
           uncertainties are dominated by the calibration error.
           [Ne$\,${\sc iii}] and [Ne$\,${\sc ii}] are ionized gas emission 
           lines (Figure~\ref{fig:spec}).
           \uline{Bottom-right:} theoretical emission of small ($r=0.35$~nm)
           neutral (red) and charged (purple) PAHs
           \citep[][for the Solar neighborhood radiation field intensity 
                    $U=1$]{draine07}.
           Large ($r=5$~nm) PAHs are shown in blue and cobalt, for comparison.
           The goal of this figure is to demonstrate the effect of the charge
           and size of the PAHs on the relative intensity of the various 
           features.
           The molecular modes of the main features are labeled in grey.}
  \label{fig:correl}
\end{figure}

In the two top panels of Figure~\ref{fig:correl}, we have highlighted in red 
the pixels corresponding to the central region.
These pixels stand out in the top right plot.
We have investigated the nature of these pixels, in the bottom-left plot of
Figure~\ref{fig:correl}, by comparing the UIB intensity to the 
[Ne$\,${\sc iii}]/[Ne$\,${\sc ii}] ratio.
The [Ne$\,${\sc iii}]/[Ne$\,${\sc ii}] ratio is known to correlate well with 
the hardness of the radiation field \citep[\textit{e.g.}][]{madden06}.
This plot shows a clear segregation between the central region and the rest of 
the galaxy.
The radiation field is harder in the center.
Although there appears to be a variation of the mean PAH size throughout the 
galaxy (as probed by the dynamics of the PAH$_{3.3}$/PAH$_{11.3}$ ratio), there 
is a hint that the central region could harbor larger PAHs, as
PAH$_{3.3}$/PAH$_{11.3}$ is lower than average.
This can be easily understood in terms of the selective destruction of the 
smallest PAHs by the hard radiation field bathing the central region.
Such a selective destruction process was previously discussed by
\citet{smith07} and \cite{sales10}, comparing integrated spectra of galaxies.
Our present data brings new evidence of this effect, within a spatially 
resolved object.

\section{SUMMARY AND PROSPECTIVES}

We have presented preliminary results of a study scrutinizing the 
variations of the UIBs in the starburst galaxy NGC$\,$1097.
The originality of our approach consists in combining {\it AKARI}/IRC and 
{\it Spitzer}/IRS spectra, in order to probe the full spectral range covered by 
UIB features.
We find evidence that the smallest PAHs are destroyed by the hard radiation 
field in the central regions.

This study is part of a larger project including $\simeq20$ nearby resolved 
galaxies.
The comparison of the UIB ratio trends between objects will provide stronger 
constraints on the origin of the processes controlling the UIB spectrum.
It might also provide constraints allowing us to favor one of the candidate 
carriers (PAH, a-C(:H), {\it etc.}).
From a methodological point of view, we plan to implement a more rigorous, 
Bayesian, spectral decomposition method, in order to better characterize the 
spread of the trends.
Finally, this study is a good test case for the JWST (launch in 2019).

\subsection*{Acknowledgments}

We ackowledge support from the PRC 1311 between France (CNRS) and Japan (JSPS).
R.W.\ and F.G.\ acknowledge support by the Agence Nationale pour la Recherche 
through the program SYMPATICO (Projet ANR-11-BS56-0023).
F.G.\ acknowledges support from the EU FP7 project DustPedia (Grant No.\ 
606847).




\bibliographystyle{$HOME/Astro/TeXstyle/Packages_AAS/aas}
\bibliography{$HOME/Astro/TeXstyle/references}

\end{document}